\documentstyle[12pt,axodraw,graphics]{article}  
\begin{document}
\pagestyle{empty}
\centerline{ \hfill          SLAC-PUB-8714
\footnote{Work supported by the Department of Energy, Contract 
DE-AC03-76SF00515}
}
\centerline{ \hfill          hep-ph/y0012258} 
\centerline{ \hfill          November 2000} 
%%%%%%%%%%<<<<<<<<<=======
%
\vskip 1cm
\begin{center}
{\bf \large
CP asymmetry in the Higgs decay into \\
the top pair due to the stop mixing}
\vskip 1cm
{Fred Browning$^{a}$, Darwin Chang$^{b,c}$, and Wai-Yee Keung$^{a}$}
\vskip 1cm
{\em $^a$Physics Department, University of Illinois at Chicago, IL
     60607-7059,  USA}\\

{\em $^b$NCTS and Physics Department, National Tsing-Hua University,\\
Hsinchu 30043, Taiwan, R.O.C.}\\

{\em $^c$Stanford Linear Accelerator Center, Stanford University, Stanford, 
     CA 94309, USA}\\

\end{center}
\date{\today}
\vskip 1.5cm
\begin{abstract}
We investigate a potentially large CP violating asymmetry in the decay
of a neutral scalar or pseudoscalar Higgs boson into the $t\bar t$
pair. The source of the CP nonconservation is  the complex mixing in
the stop $\tilde t_{L,R}$ sector.  One of the interesting consequence
is the different rates of the Higgs boson decays into CP conjugate
polarized states.

\end{abstract}
\vskip 0.5cm
%%   \centerline{PACS numbers:  14.80.Cp,  14.60.Ef,  13.40.Em}
\vskip 0.5cm
\newpage
%

%\narrowtext
\pagestyle{plain}

\section{Introduction}
The standard model (SM) of particle interactions contains one CP
violating parameter, which  is a complex phase in the quark
sector of the SM.  This phase appearing  in the quark mixing  matrix
of the charged current is expected to account for 
the observed CP violations in the $K$-$\overline{K}$ mixing, in the
$K$ decays, as well as in the potential CP violation in the
$B$-$\overline{B}$ system.

However, it is generally believed that new physics beyond the SM must
exists.  One of the major motivations for this is to understand the
seemingly unnaturalness of the Higgs mass at the electroweak scale in
the SM, the so-called gauge hierarchy problem.  In addition, due to the
difficulties of the SM to account for the baryon asymmetry of the
universe as well as to resolve the strong CP problem, it is widely
accepted that new sources of CP violation are needed.  The most
popular extension of the SM that addresses the hierarchy problem is
the supersymmetric standard model\cite{reviews,ref:higgshunt}.  The extension has
many more new (super-)particles and parameters compared to the SM.
With all these new parameters, there are many possible new sources for
CP violation.  The phenomenology of CP violation caused by these new
sources is rich and diverse.  The effect of these new sources of CP
violation may surface in the data before any super-particle is
discovered.

Even in the minimal supersymmetric standard model (MSSM), which only
augments superpartners of known particles in SM, the Higgs sector 
contains new sources of CP violation in its couplings to
super-particles.  When the $\mu$ term in the Higgs superpotential and
the soft-SUSY-breaking $A$ terms are complex, the tri-boson-couplings
between the Higgs bosons and the squarks can contain CP violation.
In MSSM with the simplest universal soft supersymmetry breaking\cite{susycp},
there are two new CP violating couplings which can be defined to be the 
phases of $\mu$ and $A$ terms in a convention that makes the others new 
couplings real.  Therefore, these new sources of CP violation are generic 
of any supersymmetric theories.  
In addition, they also have been used as one of the leading 
sources of CP violation in a scheme to use MSSM to generate baryon number 
asymmetry of the Universe in electroweak phase transition \cite{bau}.
Therefore, it should be important to look for collider phenomenology that 
can check these mechanism.
For example, these complex couplings lead to a complex phase in the 
mixing\cite{mix:stops} of stop states.  It is the purpose of this paper to 
investigate one consequence of this CP violating source in colliders.

It is expected that the future colliders are able to produce CP
violation signals\cite{ref:CK,bohdan,Soni,seoul} 
in the sectors of heavy particles.
In this article  we study the CP asymmetry in the Higgs decay
into top pairs because the large top or stop
coupling to the Higgs particles can produce largest effect.

In MSSM, even with soft breaking  terms and $R$ symmetry breaking terms,
there is no tree level mixing between the scalar and the pseudoscalar
bosons.  
Therefore their couplings can be discussed separately.
However, the scalar and the pseudoscalar bosons mix at one loop, and
their effect has to be taken into account as we will show later.

\section{Stop Mixing}
The source of CP violation that we  investigate here is due to the mixing
in the stop mass matrix.
We use the convention adopted in Ref.\cite{conv:stop}.
The mass matrix for the stop quarks in the left-right basis is given as
\begin{equation}
{\cal M}^2_{\tilde{t}} = \left(
\begin{array}{cc} m^2_{Q} + m^2_{t} + 
\Delta_{\tilde{t}_{L}} m^2_{Z} \cos 2\beta & 
-m_{t} (\mu \cot \beta + A_{t}^{*}) \\
-m_{t} (\mu^{*} \cot \beta + A_{t}) & m^2_{U} + m^2_{t} + 
\Delta_{\tilde{t}_{R}} m^2_{Z} \cos 2\beta \end{array} \right),
\end{equation}
where 
$ \Delta_{\tilde{t}_{L}}=\hbox{$1\over2$}-\frac{2}{3} \sin^2 \theta_{W}$, 
and
$ \Delta_{\tilde{t}_{R}}= -\frac{2}{3} \sin^2 \theta_{W}$.
The complex phase $\delta$ of the off-diagonal elements is the source
of   CP violation.
\begin{equation} 
\mu^{*} \cot \beta + A_{t} = |\mu^{*} \cot \beta + A_{t}| e^{i \delta}.
\end{equation}
The stop mass eigenstates, $\tilde{t}_{1}, \tilde{t}_{2}$, are related to the left and right stop states by an unitary mixing matrix
\begin{eqnarray} 
\left( \begin{array}{c} \tilde{t_{L}} \\ \tilde{t_{R}} 
\end{array} \right) 
= \left( \begin{array}{cc} 1 & 0 \\ 0 & e^{i \delta} \end{array} \right)
\left( \begin{array}{cc} \cos \theta & \sin \theta \\
-\sin \theta & \cos \theta \end{array} \right) 
\left( \begin{array}{c} \tilde{t_{1}} \\
\tilde{t_{2}} \end{array} \right)
= {\bf U} \
\left( \begin{array}{c} \tilde{t_{1}} \\
\tilde{t_{2}} \end{array} \right)
\end{eqnarray}
%This then reduces to a single mixing matrix ${\bf U}$, 
%\begin{eqnarray} 
%\left( \begin{array}{c} \tilde{t_{L}} \\ \tilde{t_{R}}
%\end{array}\right) 
%={\bf U} \left(\begin{array}{c}\tilde{t_{L}}\\ \tilde{t_{R}}\end{array}\right)
%=
%\left( \begin{array}{cc} \cos \theta & \sin \theta \\
%-\sin \theta \, e^{i \delta} & \cos \theta \, e^{i \delta} \end{array}%\right)
% \left( \begin{array}{c} \tilde{t_{1}} \\
%\tilde{t_{2}} \end{array} \right). 
%\end{eqnarray}
The masses of these eigenstates are given by
\begin{equation}
m^2_{1,2}=\hbox{$1\over2$} 
(m^2_{Q} + m^2_{U}  + 2 m_{t}^2 
 + (\hbox{$1\over2$}-\hbox{$4\over3$} \sin^2\theta_W) 
                  m_{Z}^2 \cos 2 \beta \mp \sqrt{R} ) \ ,
\end{equation}
\begin{equation} 
R=\left( m^2_{Q} -m^2_{U} + 
{1\over2} m_Z^2\cos2\beta\right)^2
+ 4 m_{t}^2 |\mu \cot \beta + A^{*}_{t}|^2   \ .
\end{equation}
Here we denote   $\tilde{t}_{1}$ as the lighter state.
The mixing angle is given as
\begin{equation} 
\tan \theta = -\frac{[m_{Q}^2 - m_{U}^2 + \hbox{$1\over2$}
 m_{z}^2 \cos 2\beta +
\sqrt{R}]}{2 m_{t} |\mu \cot \beta + A_{t}^{*}|}.
\end{equation}
Strong gluino couplings to stops and tops in the left-right basis
is given by 
\begin{equation} 
{\cal L}_{\tilde g}
 = - \sqrt{2} g_{s} \overline{t_{R}} \tilde{g} \tilde{t_{R}}
- \sqrt{2} g_{s} \overline{t_{L}} \tilde{g} \tilde{t_{L}} \ .
\end{equation} 
In terms of mass eigenstates,
\begin{equation} {\cal L}_{\tilde g}         
= -\sqrt{2} g_{s} \overline{t}(P_{R} \cos \theta - P_{L} \sin
\theta e^{i \delta})  \tilde{t_{1}} \tilde{g} - \sqrt{2} g_{s} \overline{t} ( P_{R} \sin
\theta + P_{L} \cos \theta e^{i \delta}) \tilde{t_{2}} \tilde{g},
\end{equation}
where $P_{L}$ is the left projection 
$\hbox{$1\over2$}(1-\gamma^{5})$, and $P_{R}$ is
the right projection $ \hbox{$1\over2$}(1+\gamma^{5})$.
We also need the stop-stop coupling to $Z$,
\begin{equation} 
{\cal L}_Z= -{g\over \cos \theta_{W}} \left[ 
(\hbox{$1\over2$} -\hbox{$2\over3$}\sin^2\theta_W) 
\tilde{t}_{L}^{\dagger}\stackrel{\leftrightarrow}{i\partial^\mu}\tilde{t}_{L}
-\hbox{$2\over3$}\sin^2\theta_W
\tilde{t}_{R}^{\dagger}\stackrel{\leftrightarrow}{i\partial^\mu}\tilde{t}_{R}
                                    \right]Z_{\mu} \ .
\end{equation}
After mixing, the Lagrangian for the $Z$ coupling 
in the $ \tilde{t}_{1}$, $\tilde{t}_{2}$ basis is
\begin{equation}
{\cal L}_{Z}  =
-\frac{g}{\cos \theta_{W}} \left[
 \left( \hbox{$1\over2$}\cos^2 \theta
- \hbox{$2\over3$} \sin^2\theta_{W}\right)    
\tilde{t}_{1}^{\dagger} \stackrel{\leftrightarrow}{i\partial^\mu}\tilde{t}_{1}
+ \left( \hbox{$1\over2$} \sin^2\theta
   - \hbox{$2\over3$} \sin^2\theta_{W}\right)
\tilde{t}_{2}^{\dagger}\stackrel{\leftrightarrow}{i\partial^\mu}\tilde{t}_{2}
 \right. \end{equation}
$$  \qquad\qquad \left.
+\hbox{$1\over4$}\sin(2\theta)
\left( \tilde t_1^\dagger \stackrel{\leftrightarrow}{i\partial^\mu} \tilde t_2
      +\tilde t_2^\dagger \stackrel{\leftrightarrow}{i\partial^\mu} \tilde t_1 \right)
\right]Z_{\mu} \ . $$
Note that the last term is real in this phase convention.

\section{The Higgs couplings to stops}
In MSSM, there is only one pseudoscalar boson., $A^0$.  The pseudoscalar Higgs coupling to the stop squarks is given by the Lagrangian
\begin{equation} {\cal L}_{A}= 
\left( \tilde{t}_{1}^\dagger \; \tilde{t}_{2}^\dagger \right)
{\bf T}^{A}
\left( \begin{array}{cc} \tilde{t}_{1} \\ \tilde{t}_{2} \end{array} \right).
A^0
\end{equation}
The matrix $ {\bf T}^{A} $ is given as
\begin{equation}
{\bf T}^{A}  = \frac{m_{t}}{v_{2}}
\left( \begin{array}{cc} 2 \sin \theta \cos \theta \; {\rm Im} (\hat{A} )&
-i( \cos^2 \theta \hat{A}^{*} + \sin^2 \theta \hat{A}) \\
i( \cos^2 \theta \hat{A} + \sin^2 \theta \hat{A}^{*}) &
-2 \sin \theta \cos \theta \; {\rm Im} (\hat{A} )\end{array} \right), 
\end{equation}
and $\hat{A}$ is defined as 
$\hat{A} =(A_{t} \cos \beta - \mu^{*} \sin \beta)e^{-i \delta}$.
Note that the nonvanishing of $T^A_{11}$ or $T^A_{22}$ is a sure sign of CP 
violation already (similar to $K_L \rightarrow 2 \pi$).  
However, if for some reason $\mu$ and $A_t$ happen to have the same phase, 
$T^A_{11}$ and $T^A_{22}$ will vanish because in this very special case the 
phase in the stop mass matrix and that in the pseudoscalar couplings can be 
removed simultaneously.

The pseudoscalar Higgs coupling to the top quark is given by the following
Lagrangian,
\begin{equation}
{\cal L}^Y_{A}
=   {g m_{t}\over 2 m_W} 
     \cot \beta \; \overline{t} i \gamma^{5} t A^0.
\end{equation}

The neutral scalar Higgs sector is made up two scalar eigenstates, 
$H^{0}$ and $h^{0}$.
There masses are given as
\begin{equation} 
m^2_{H,h} = \hbox{$1\over2$} \left[ m^2_{A} + m^2_{Z} \pm 
\sqrt{(m^2_{A} + m^2_{Z})^2 - 4 m^2_{A} m^2_{Z} \cos 2 \beta}\right].
\label{eq:treemass}
\end{equation} 
Since in MSSM the constraint on the lightest scalar, $h$, is such that it is 
too light to decay into the top pair, we shall concentrate on the decay of the 
heavy Higgs boson, $H$, which can decay into the top pair.   
Our general framework can also be used for the decay of the lighter
boson, $h$, of course, if for any reason that it should be heavy enough.
The heavy Higgs coupling to the stops in the left-right basis is given as
\begin{equation} 
{\bf T}_0^{H}= \left( \begin{array}{cc} 
-\frac{g m_{Z}}{\cos \theta_{W}}
\Delta_{\tilde t_L} \cos(\alpha+\beta) - \frac{g m_{t}^2 \sin \alpha}{m_{W}
\sin \beta}  & 
\frac{g m_{t}}{2 \sin \beta} (A_{t}^{*} \sin \alpha + \mu \cos \alpha)
\\  \frac{g m_{t}}{2 \sin \beta} (A_{t} \sin \alpha + \mu^{*} \cos \alpha) 
& -\frac{g m_{Z}}{\cos \theta_{W}}
\Delta_{\tilde t_R} \cos(\alpha+\beta) - 
\frac{g m_{t}^2 \sin \alpha}{m_W\sin \beta} 
\end{array} \right), \end{equation}
where the mixing angle $\alpha$ is given in Ref. \cite{ref:higgshunt}.
This matrix must then be transformed into the stop mass eigenstates.  
This is accomplished by using the stop mixing matrix,
\begin{equation} 
 {\bf T}^{H}= {\bf U}^{\dagger} {\bf T}_0^{H} {\bf U}. 
\end{equation}
Its Yukawa coupling is 
\begin{equation} {\cal L}^Y_{H}
 = \frac{g m_{t} \sin \alpha}{2 m_W \sin \beta} 
\;\overline{t} t H^{0}.\end{equation}

\section{Helicity calculation of the matrix element}

To get non-zero CP asymmetry in Higgs decays, 
in addition to CP violating  couplings, 
it is necessary to get the absorptive parts from the decay 
amplitudes in order to overcome the constraint from the $CPT$ theorem.
We labeled $S^I$ and  $P^I$ as the absorptive form factors 
of the Higgs or pseudo-Higgs couplings to the top quark.
They begin to appear at the 1-loop level, unlike 
their  dispersive parts $S$ and $P$, which can exist at the tree level,
\begin{eqnarray} 
{\cal M} = \overline{u}(p) \left[ (S+iS^I) {\bf 1} + i (P+iP^I) \gamma ^{5}
                           \right]
	v(p')\ .		 \end{eqnarray}
In the Weyl representation, the $\gamma$ matrices are given by
\begin{eqnarray*} \gamma^{5}= 
\left( \begin{array}{cc} -1 & 0 \\ 0 & 1 \end{array} 
\right) & \gamma^{0}= \left( \begin{array}{cc} 0 & 1 \\ 1 & 0 \end{array} 
 	\right).
 \end{eqnarray*}
The free spinors of momenta $p,p'$ and helicities 
$\lambda, \lambda'$ are given by
\begin{eqnarray} 
u(p,\lambda)= \left( \begin{array}{c}  \omega _{- \lambda} \chi_{+ \lambda} \\
 	  \omega_{+\lambda} \chi _{+\lambda} \end{array} \right) \ ,
   \qquad 
v(p',\lambda)= 
\left( \begin{array}{c} -\lambda' \omega _{+\lambda'} \chi_{-\lambda'}\\
	\lambda' \omega_{-\lambda'} \chi _{-\lambda'} \end{array} \right), 
\nonumber \end{eqnarray}
where the $\chi$'s are  two component spinor eigenfunctions
$\vec{\sigma} \cdot \hat{p} \; \chi_{\lambda} (p) 
= \lambda \; \chi_{\lambda} $.
The $\omega_{\pm}$ are functions 
of the energy and momentum of the particles, 
$\omega_{\pm}=\sqrt{E\pm |p|}$.
Notice that the helicities of $t\bar t$ must match
$\lambda'=\lambda$ because  of conservation of angular momentum.   
Our normalization of the  spinor is 
$u^{\dagger}_{\lambda} u_{\lambda}  = v^{\dagger}_{\lambda} v_{\lambda}=2E$. 
The asymmetry between the left and right matrix elements is given by
\begin{equation} {\cal A}= \frac{|{\cal M}_{LL}|^2-|{\cal M}_{RR}|^2}
	{|{\cal M}_{LL}|^2+|{\cal M}_{RR}|^2}.
\label{eq:assy}
\end{equation}
The matrix elements are given by
\begin{eqnarray}
{\cal M}_{LL} &=& \sqrt{s} [- \beta_t (S+iS^I) -i(P+iP^I)]\ ,\\
{\cal M}_{RR} &=& \sqrt{s} [- \beta_t (S+iS^I) +i(P+iP^I)]\ ,
\end{eqnarray}
with $\beta_t=(1-4m_t/s)^{1\over2}$ and $s=m_H^2$.
The asymmetry can finally be obtained using the definition from
Eq.~(\ref{eq:assy}),
\begin{equation} {\cal A}= 
\frac{2 \beta_{t} (PS^{I}-P^{I}S)}{P^{I2}+P^2+\beta_{t}^2S^2
+\beta_{t}^2 S^{I2}}    \ . \label{eq:assy2} 
\end{equation}
Since we assume the Higgs boson has definite CP parity at the tree
level, the final state interactions due to exchanging gluons or gauge
bosons in Ref.~\cite{ref:CK} are not able to produce this CP asymmetry
at the one-loop level. However, the rich CP phases in the sector of
SUSY partners, especially the gluino and the stop, can give rise to
large ${\cal A}$.  For scalar boson decay, the second term $P^{I}S$ in 
${\cal A}$ gives the leading contribution; while for pseudoscalar boson decay, 
the first term, $PS^{I}$ is the leading contribution.

The polarization asymmetry is Eq.(\ref{eq:assy2}) can be translated
into the lepton energy asymmetry\cite{ref:CK,ref:Peskin,ref:kly}
in the final semileptonic channel $t\to b \ell^+\nu$. The
energy $E_0(\ell^+)$ distribution of a static $t$ quark decay
$t\rightarrow \ell^+\nu b$ 
is very simple in the narrow width $\Gamma_W$
approximation when $m_b$ is negligible.
\begin{equation}
    f(x_0)= \left\{
              \begin{array}{lll}
       x_0 (1-x_0)/D & \quad\  & \mbox{if $m_W^2/m_t^2 < x < 1$}, \\
       0             & \quad\  & \mbox{otherwise.}
             \end{array}
              \right.
\end{equation}
Here we denote the scaling variable $x_0=2E_0(\ell^+)/m_t$ and
the normalization factor $D={1\over 6}-{1\over 2}(m_W/m_t)^4
+{1\over 3}(m_W/m_t)^6$.
When the $t$ quark is not static, but moves at a speed $\beta_t$
with helicity $L$ or $R$, the distribution expression becomes
a convolution,
\begin{equation}
     f_{R,L}(x,\beta_t)=
    \int_{x/(1+\beta_t)}^{x/(1-\beta_t)} f(x_0)
            {\beta_t x_0 \pm (x-x_0) \over 2 x_0^2\beta_t^2}
      dx_0
\;.
\end{equation}
Here $x=2E(\ell^+)/E_t$. The kernel above is related to the
$(1\pm\cos\psi)$  polar angular distribution.
Similar distributions for the $\bar t$ decay is related by CP
conjugation at the tree--level.
Using the polarization asymmetry formula in  Eq.(\ref{eq:assy2}),
we can derive expressions for the energy distributions of
$\ell^-$ and $\ell^+$:
\begin{equation}
N^{-1}dN/dx(l^\pm)
=\hbox{$1\over2$}(1\pm{\cal A})f_L(x,\beta_t)+
\hbox{$1\over2$}(1\mp{\cal A})f_R(x,\beta_t)
\;.
\label{eq:dist}
\end{equation}
Here distributions are compared at the same energy for the lepton
and the anti--lepton at the rest frame of the Higgs boson, 
$x(\ell^-)=x(\ell^+)=x=4E(\ell^\pm)/M_H$.
To prepare a large sample for analysis, we only require that each
event
has at least one prompt anti--lepton $\ell^+$
from the $t$ decay {\it or}
one prompt lepton $\ell^-$ from the $\bar t$ decay.

\section{Absorptive parts of 3-point vertices}
We first study  the triangle diagram via gluino exchange
in Fig.~1..
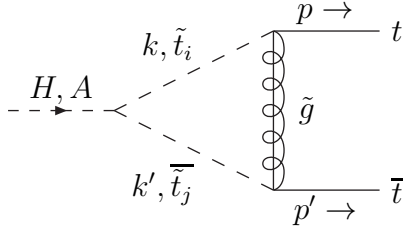
\begin{figure}
\begin{center}
\begin{picture}(160,80)(5,5)
\Text(120,75)[b]{$p \to$}  \Text(120,8)[t]{$p' \to$}
\Line(100,70)(140,70)   \Text(145,70)[l]{$t$}
\Line(100,10)(140,10)   \Text(145,10)[l]{$\overline{t}$}
\DashLine(40,40)(100,70){5}  \Text(70,65)[r]{$k,\tilde t_i$}
\DashLine(40,40)(100,10){5}  \Text(70,20)[rt]{$k',\overline{\tilde t_j}$}
\Line(100,70)(100,10)
\Gluon(100,70)(100,10){4}{5} \Text(110,40)[l]{$\tilde g$}
\DashArrowLine(0,40)(40,40){5} \Text(20,45)[b]{$H,A$}
\end{picture}
\caption{Triangle diagram via gluino exchange}
\end{center}
\end{figure}

\subsection{The ${\cal M}_{11}$ stop loop}

When the intermediate state is 
$\tilde{t}_{1} \overline{\tilde{t}_{1}}$, the Feynman rule gives
\begin{eqnarray}  
i {\cal M}_{11} & = & (-i \sqrt{2} g_{s} )^2 \int {\cal N}_{11}
\frac {i} 
{k^2-m_{1}^2}  
\frac{i}{k'^2-m_{1}^2} \frac{d^{4}q}{(2 \pi)^{4}}  
iT_{11} C_{F}, 
\end{eqnarray} 
where $ {\cal N}_{11}$ is defined as 
\begin{equation} {\cal N}_{11} = \overline{u}(p)  
(P_{R} \cos \theta - P_{L} \sin \theta e^{i \delta})  
\frac{i(\not\!q  +  m_{\tilde{g}})}{q^2-m_{ \tilde{g}}^2} 
 (P_{L} \cos \theta - P_{R} \sin \theta e^{-i \delta}) v(p'). 
\end{equation} 
The color factor $C_{F}$  is $\frac{4}{3}$.
The absorptive part of the amplitude which is
needed for CP violation is obtained by cutting 
across the momentums $k$ and $k'$.  
\def\dum{
First we replace propagators that the cut is  made
on with the on-shell conditions. Taking the momentum $k$ as
an example one gets
\begin{equation} 
\int \frac{1}{k^2-m_{1}^2} \rightarrow 
\int (-2 \pi i)\delta (k^2-m_{1}^2).
\end{equation}
The same must be done for the other momentum $k'$.  
Now the integral is changed from $q$, the 
transfer momentum, to the momenta $k$ and $k'$.  
This is accomplished by changing the 4
dimension integration from $q$ to $k$ and $k'$.  
Also a delta function is needed to make sure
that the conservation of energy is upheld,
\begin{equation} 
\frac{d^{4}q}{(2 \pi)^{4}} \rightarrow \frac{d^{4}k}{(2 \pi)^{4}} 
\frac{d^{4}k'}{(2 \pi)^{4}} (2 \pi)^{4} \delta^{4} (k+k'-p-p') \ .
\end{equation}
}
%%%%
The discontinuity\cite{cut-rule} of the matrix element is
$$ {\rm Disc}(i {\cal M}_{11}) =
\frac{g_{s}^2 T_{11}}{8\pi} 
\beta_{1} C_{F}   $$
\begin{equation}
\times \int  \overline{u}(p) 
{ \not\!q (1-\gamma_5\cos2\theta)+ m_{\tilde{g}} \sin(2 \theta)
(-\cos \delta + i\gamma^5 \sin \delta) 
\over q^2-m_{\tilde g}^2 } v(p')   {d \Omega_{k} \over 4 \pi} \ .
\end{equation}
The phase space integration involves the following forms,
\begin{equation}
J_{ij}\equiv
 s\int \frac{1}{q^2-m_{\tilde{g}}^2} \frac{d \, \Omega_{k}}{4 \pi}. 
= \frac{1}{\beta_{t} \beta_{ij}} \ln \left(
\frac{
\beta_{t}^2 + \beta_{ij}^2 -2 \beta_{t} \beta_{ij} +4 m_{\tilde{g}}^2/s}
{
\beta_{t}^2 + \beta_{ij}^2 + 2 \beta_{t} \beta_{ij} +4 m_{\tilde{g}}^2/s}
\right) ,\end{equation}
\begin{equation} s \int \frac{q^{\mu}}{q^2-m_{\tilde{g}}^2} 
\frac{d \Omega_{k}}{4 \pi} = -H_{ij} (p-p')^{\mu} + K_{ij}  (p+p')^{\mu}. 
\label{eq:HK}
\end{equation}
Multiplying both sides by $(p-p')_{\mu}$ the $H_{ij}$ function 
can be isolated out because $(p+p')\cdot (p-p')$ is zero.
The $H_{ij}$ function for any intermediate mass $ m_{i}$ and $m_{j}$ is
\begin{equation}
 \beta_t^2 H_{ij} =  1 + \hbox{$1\over4$}(\beta_{ij}^2 
     - \beta_{t}^2 + 4 m_{\tilde{g}}^2/s)J_{ij}  \ .
\end{equation}
The $ \beta_{ij}$ function is given by
\begin{equation} \beta_{ij}  = \sqrt{1-2(m_{i}^2+m_{j}^2)/s + 
(m_{i}^2-m_{j}^2)^2/s^2}.
\end{equation}
Notice that when $ i=j $, $\beta_{ij}$ reduces to 
$\beta_{i}= \sqrt{1-4m_{i}^2/s}$.
The function $K_{ij}$  is obtained by 
contracting  Eq.~(\ref{eq:HK}) with $p+p'$.
\begin{equation} K_{ij}=-\hbox{$1\over2$} J_{ij}(m_{i}^2-m_{j}^2)/s 
\ . \end{equation}
Notice that for matrix elements with both the stop and the anti-stop 
of the same type, the term proportional to $(p+p')$ in Eq.~(\ref{eq:HK}) 
does not contribute.

After this, the imaginary part of the matrix element is needed.  The
imaginary parts are obtained by using the relation
\begin{equation} 
 {\rm Disc}({\cal M})= 2 i
\overline{u}(p) \left[ S^{I} {\bf 1} + P ^{I} i\gamma ^{5} \right]
	v(p') \ .
\end{equation}
\begin{eqnarray}  
S^{I}_{11} & = &  \frac{g_{s}^2 T_{11} \beta_{1}}{16 \pi s} C_{F}
\left(
+m_{\tilde{g}}J_{11}\sin(2\theta)\cos\delta+2m_t H_{11}
\right)  \ , \nonumber\\       
P^{I}_{11} & = &  \frac{g_{s}^2 T_{11} \beta_{1}}{16 \pi s} C_{F}
\left(-m_{\tilde{g}}J_{11}\sin(2\theta)\sin\delta  \right)    \ .
\label{eq:oneone}
\end{eqnarray}

\subsection{The ${\cal M}_{22}$ stop loop}
Similarly, we obtain results  for the intermediate state 
$\tilde{t}_{2} \overline{\tilde{t}_{2}}$.
The matrix element is given by
\begin{eqnarray} i {\cal M}_{22} 
& = & (-i \sqrt{2} g_{s} )^2 \int {\cal N}_{22} \frac {i}
{k^2-m_{2}^2} \frac{i}{k'^2-m_{2}^2} \frac{d^{4}q}{(2 \pi)^{4}} iT_{22} C_{F},
\end{eqnarray}
where $ {\cal N}_{22}$ is given by
\begin{equation} {\cal N}_{22} = \overline{u}(p) 
(P_{R} \sin \theta + P_{L} \cos \theta e^{i \delta}) 
\frac{i(\not\!q  + m_{\tilde{g}})}{q^2-m_{ \tilde{g}}^2}
 (P_{L} \sin \theta + P_{R} \cos \theta e^{-i \delta}) v(p').
\end{equation}
After integrating the phase space of the intermediate state in the cut
diagram,  the  form factors are
\begin{eqnarray}  S^{I}_{22}  & = & 
\frac{g_{s}^2 T_{22} \beta_2}{16 \pi s} C_{F} \left(
-m_{\tilde g}J_{22}\sin(2\theta)\cos\delta+2m_t H_{22}
\right) \ ,\\
    P^{I}_{22} & = &  \frac{g_{s}^2 T_{22} \beta_2}{16 \pi s} C_{F} \left(
+m_{\tilde g}J_{22}\sin(2\theta)\sin\delta \right)  \ . \nonumber
\label{eq:twotwo}
\end{eqnarray}

\subsection{The ${\cal M}_{12}$  stop loop}

The amplitude involving 
the intermediate state  $\tilde{t}_{1} \overline{\tilde{t}_{2}}$
is given by 
\begin{eqnarray} i {\cal M}_{12} & = & 2 i^{6} g_{s}^2T_{12} C_{F}
\int {\cal N}_{12} \frac{1}{k^2-m_{1}^2} \frac{1}{k'^2-m_{2}^2}
\frac{d^{4}q}{(2 \pi)^{4}}  \nonumber\\
{\cal N}_{12} & = & \overline{u}(p) (P_{R} \cos \theta - 
P_{L} \sin \theta e^{i \delta}) \frac{(\not\!q  + 
m_{\tilde{g}})}{q^2-m_{ \tilde{g}}^2} (P_{L} \sin \theta + P_{R} \cos \theta 
e^{-i \delta}) v(p'). 
\end{eqnarray}
For $\tilde{t}_{2} \overline{\tilde{t}_{1}}$, it is
\begin{eqnarray} i {\cal M}_{21} & = & 2 i^{6} g_{s}^2T_{21} C_{F}
\int {\cal N}_{21} \frac{1}{k^2-m_{2}^2} \frac{1}{k'^2-m_{1}^2}
\frac{d^{4}q}{(2 \pi)^{4}}     \ , \\
{\cal N}_{21} & = & \overline{u}(p) (P_{R} \sin \theta + 
P_{L} \cos \theta e^{i \delta}) \frac{(\not\!q + 
m_{\tilde{g}})}{q^2-m_{ \tilde{g}}^2} (P_{L} \cos \theta - P_{R} \sin \theta 
e^{-i \delta}) v(p')\ . \nonumber
\end{eqnarray}
After integrating over the intermediate phase space,
we add up the absorptive parts to give
\begin{eqnarray} 
S^{I}_{12+21}&=&
-\frac{g_{s}^2 \beta_{12}}{8 \pi s} C_{F} 
 {\rm Re}\left[m_{\tilde g} T_{21}(\cos^2\theta e^{i\delta} 
            -\sin^2 \theta e^{-i \delta})\right]J_{12}     \ , \\
P^{I}_{12+21}&=& -\frac{g_{s}^2  \beta_{12}}{8 \pi s} C_{F} 
 {\rm Im}\left[-m_{\tilde g}T_{21}(\sin^2\theta e^{-i\delta} 
            +\cos^2 \theta e^{i \delta}) \right. \nonumber\\ &&
\qquad \qquad \left. 
 + m_t\sin( 2 \theta) T_{21}(m_2^2-m_1^2)/ s\right]J_{12}
\ .
\end{eqnarray}
\section{Absorptive parts of 2-point vertices}
We study the bubble loops which involve only the $\overline{\tilde
t}\tilde t$  pair.
\subsection{$Z$ diagrams}
\begin{figure}
\begin{center}
\begin{picture}(65,50)(0,-40)
\Text(-8,-1)[t]{$l=p+p'$}
\DashLine(0,0)(20,0){4}      \Text(0,8)[b]{$H\to$}
\DashCArc(30,0)(10,180,0){4}  \Text(30,15)[b]{$\tilde t_i$}
\DashCArc(30,0)(10,0,-180){4} \Text(30,-15)[t]{$\tilde t_j$}
\Photon(40,0)(60,0){4}{3}    \Text(50,8)[b]{$Z$}
\Line(60,0)(90,30)           \Text(95,30)[l]{$t,p$}
\Line(60,0)(90,-30)          \Text(95,-30)[l]{$\overline t,p'$}
\end{picture}
\end{center}
\caption{$Z$ exchange diagram}
\end{figure}
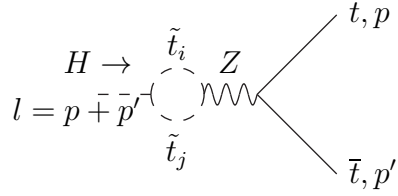
$Z$ diagrams that contain $\bar{\tilde{t}_{1}} \tilde{t}_{1}$ or
$\bar{\tilde{t}_2} \tilde{t}_{2}$ are identical zero because of the phase
space integration. This point  will become obvious from the result of 
the mixed intermediate states $\bar{\tilde{t}}_{1} \tilde{t}_{2}$ or
$\bar{\tilde{t}}_{2} \tilde{t}_{1}$.
The $ {\cal M}_{12}$ matrix element is given below,
\begin{eqnarray}
i {\cal M}_{12} & = & 
\frac{g^2 T_{12} N_{C}}{4 \cos^2 \theta_{W}} \sin( 2 \theta)
 \int {\cal N}_Z \frac{1}{l^2 - m_{Z}^2} 
\frac{1}{ k^2-m_{1}^2}
\frac{1}{{k'}^2-m_{2}^2} \frac{d^{4}k}{(2 \pi)^{4}}, \nonumber
\end{eqnarray}
where $ {\cal N}_Z$ is given as 
\begin{equation} {\cal N}_Z = \overline{u}(p) \gamma^{\mu} 
\left( 
\hbox{$1\over4$} - \hbox{$2\over3$} \sin^2 \theta_{W} 
- \hbox{$1\over4$} \gamma^{5}
\right) v(p') 
\left(g_{\mu\nu} -l_\mu l_\nu/ m_Z^2 \right) (k-k')_{\nu} \ .
\end{equation}
The $ {\cal M}_{21}$ matrix element is 
very similar to the above matrix element,
with the substitution of $T_{12}$ by $T_{12}^{*}$,
\begin{equation} 
P^{I}_{12+21;Z} = 
\frac{g^2 N_{C}}{64 \pi} \frac{m_{t} \beta_{12} \sin (2 \theta)}
{ m_{Z}^2\cos^2 \theta_{W} } \frac{m_{1}^2-m_{2}^2}{s} {\rm Im}(T_{12})
\ ,
\end{equation}
where the color factor $N_{C} =3$.  This graph will contribute only to 
 CP  violation of the scalar Higgs decay.  
One may think that without gluino couplings in the graph, one 
should be able to rotate away the CP violating phase in scalar coupling
$T^H_{12}$.  However, such rotation will produce a complex phase in 
$\tilde t_1^\dagger \tilde t_2 Z$ coupling in Eq.(10).

\subsection{$A^0$--$H^0$ Mixing}

The stop bubble loop induces $A^0$--$H^0$ Mixing. We study its 
absorptive part which contributes to the  CP  violation.
In the heavy Higgs mass limit of MSSM, $m_{A^0}$ and $m_{H^0}$ are quite
close to each other based on the tree-level mass relation
in Eq.(\ref{eq:treemass}).  
However,  it is known that there is large higher order correction to
the tree-level mass relation. Thus in our
following study we allow the masses $m_{A^0}$ and $m_{H^0}$ to vary
independently, not restricted by the tree-level formula.
\begin{figure}
\begin{center}
\begin{picture}(65,60)(0,-30)
\DashLine(0,0)(20,0){4}      \Text(10,8)[b]{$A^0$}
\DashCArc(30,0)(10,180,0){4}  \Text(30,15)[b]{$\tilde t_i$}
\DashCArc(30,0)(10,0,-180){4} \Text(30,-15)[t]{$\tilde t_j$}
\Photon(40,0)(60,0){4}{3}    \Text(50,8)[b]{$H^0$}
\Line(60,0)(90,30)           \Text(95,30)[l]{$t$}
\Line(60,0)(90,-30)          \Text(95,-30)[l]{$\overline t$}
\end{picture}
\end{center}
\caption{Higgs mixing diagram}
\end{figure}
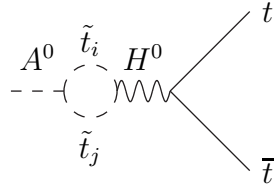
The matrix element for the pseudoscalar Higgs turning into a stop pair,
then becoming the heavy Higgs, and finally decaying into a top pair,
is given as
\begin{equation}i
{\cal M}=  \frac{i g m_{t} \sin \alpha}{2 m_{W} \sin \beta} 
    \overline{u}(p) v(p') 
\sum_{ij} \int \frac{T_{ji}^{H^{0}}}{m_{A^0}^2-m_{H}^{2}} 
     \frac{T_{ij}^{A^{0}}}{(\frac{l}{2}+q)^{2}-m_{i}^{2}} 
     \frac{id^{4}q/(2 \pi)^{4}}{(\frac{l}{2}-q)^{2}-m_{j}^{2}}   \ .
\end{equation}
Making the same cut as that in the $Z$ loop digram, we obtain
the imaginary part of the form factor,
\begin{equation}
S^I(A^0\to H^0\to\bar t t) = -
\frac{g m_{t}}{32 \pi m_{W}} {\sin \alpha \over \sin \beta}
\sum_{ij}\beta_{ij}
\frac{T^{H^{0}}_{ji} T^{A^{0}}_{ij}}{m_{A^0}^2  - m_{H^{0}}^{2}}  \ .
\end{equation}
A similar expression is derived  for 
the heavy scalar Higgs decay,
\begin{equation}
P^I(H^{0}\to A^0\to \bar t t) = -
\frac{g m_{t}}{32 \pi m_{W}} \cot\beta
\sum_{ij}  \beta_{ij}
\frac{T^{H^{0}}_{ji} T^{A^{0}}_{ij}}{ m_{H^0}^2-m_{A^{0}}^{2}} \ .
\end{equation}

\section{Physical and Numerical Analyses}

Before we plunge into the numerical analysis, 
it is interesting to check the limit
in which the two stop states are accidentally degenerate.   In that case, 
$({\cal M}^2_{\tilde{t}})_{12} = ({\cal M}^2_{\tilde{t}})_{21} = 0$,
$({\cal M}^2_{\tilde{t}})_{11} = ({\cal M}^2_{\tilde{t}})_{22}$.  
Therefore $\mu^{*} \cot \beta + A_{t}=0$, 
and $\mu^{*}$ and $A_{t}$ should have 
the same phase which can still serve as the source of CP violation.  
In that case, $\theta$ and 
$\delta$ in ${\bf U}$ in Eqs.(3),(12),(16) and in the definition of $\hat{A}$, 
should be set to zero.  

Thus in this limit, the stop loops 
do not contribute to 
${\cal M}_{11}$ and ${\cal M}_{22}$ 
in the pseudoscalar case, because 
${T}^{A}_{11} = {T}^{A}_{22} = 0$.
However they still give rise to 
 CP  violation in ${\cal M}_{12}$, ${\cal M}_{21}$ 
because 
${T}^{A}_{12} = (-i m_t/v_2)(A^{*}_{t} \cos \beta - \mu\sin\beta)$.  
One may attempt to absorb this phase by rotating the phase of, say, the
right stop, however such rotation will lead to complex gluino-top-stop
couplings which cannot be rotated away because of the nonvanishing
gluino mass.  From this, it is easy to understand why a factor of
gluino mass has to appear in Eq.(41) for $S^{I}_{12+21}$.  Similarly,
for the scalar Higgs decay in the degenerate stop limit, 
the stop loops still produce no  CP  violating effect in
${\cal	M}_{11}$	and	${\cal	M}_{22}$,
because $\sin\theta = 0$
and only the term proportional to the gluino mass in $P^{I}_{12+21}$
contributes  as reflected in Eq.(42).

It is also straightforward 
to note that in the degenerate limit, the contributions of both 
$H^0$--$Z^0$ and $A^0$--$H^0$ bubble graphs vanish.  
In the $H^0$--$Z^0$ case, the phase of the scalar Higgs, $H$, 
coupling as well as that of the stop mixing can be rotated away simultaneously 
(into the gluino couplings) 
without affecting the $Z$ coupling and this is reflected 
in $m^2_1 - m^2_2 =0$ factor in Eq.(44). 
For $A^0$--$H^0$, 
the phase of pseudoscalar coupling as well as that of the 
scalar coupling can be rotated away simultaneously 
also and this is reflected in 
$$ \sum_{ij}\beta_{ij}T^{H^{0}}_{ji} T^{A^{0}}_{ij} = 
\beta_{12}(T^{H^{0}}_{21} T^{A^{0}}_{12} 
         + T^{H^{0}}_{21} T^{A^{0}}_{12}) = 0 \ ,$$  
in this particular limit.
%%%%%%%%%%%%%

To illustrate our result numerically, in the following, we set the
parameters so that only the lighter stops states $\overline{\tilde
t_1} \tilde t_1$ are light enough to be on-shell for simplicity.  In
such a scenario, only some of the above contributions are available.
In Fig.~4, we show the mass $m_1$ of the lightest stop versus
$\tan\beta$.  The best current limit of the lowest bound on lightest
stop mass from LEP is about 95 GeV
\cite{ref:lepstopbound}, and this means that $\tan \beta < 3$ is
not allowed for the case $m_{Q}=m_{U}=300$GeV,
$\mu=500\,$GeV, $A_{t}=500 e^{\frac{i \pi}{4}}\,$GeV.
If one wishes to study the
possibility of a much heavier Higgs which can decay to all channels of
stops, the remaining diagrams can be easily incorporated into the
numerical analysis.

\subsection{Pseudoscalar-Higgs Decay}

In the model of our study, the $A^0$ remains its status as a
pseudoscalar boson at the tree level. 
CP violation in the pseudoscalar Higgs decay into top pairs occurs 
starting at the one-loop level.  
The leading contribution requires 
induced scalar form factor $S^I$ which, as we have shown, can be obtained 
from the absorptive part due to the intermediate 
$\overline{\tilde t}\tilde t$ state.
Notice that there is no $Z$ loop contribution to $S^I$ in 
the Higgs decay.
Fig.~5 show the asymmetry for the pseudoscalar
Higgs decay as defined by (\ref{eq:assy2}).  Fig.~6 shows
the branching ratios of the pseudoscalar-Higgs decay to top pairs, bottom
pairs, and stop pairs.  For small $ \tan \beta$ the decay channel is
mostly top pairs.

\subsection{Higgs Decay}

In the Higgs decay, the CP violation is caused by terms proportional
to the $P^I$ form factors.  The $Z$ diagrams can contribute 
in principle if not disallowed by the kinematics. 
As stated before it does not contribute in our illustration because 
we assume a heavy $\tilde{t}_{2}$.
Fig.~7 shows the CP asymmetry of the Higgs decay.  The
branching ratios for the Higgs to decay into tops, bottoms, stops,
$W$'s, and $Z$'s are given in Fig.~8.  
%%
%% Again small $\tan
%% \beta$ is preferred for the Higgs decay into top pairs.

\section{Conclusion}
The complex mixing among the stop sector can produce CP asymmetry at the
level of a few percent in the final products of polarized $t\bar t$
states from the Higgs boson decay.  Such asymmetry can be measured in
the energy spectra of the final leptons. Unlike the usual
two-Higgs-doublet model, the CP violation does not require the mixing 
among   $A^0$ and $H^0$ states at the tree  level.  

This work was supported in parts by National Science Council of
R.O.C., and by U.S. Department of Energy (Grant No. DE-FG02-84ER40173).
%

%%%%%%%%%%%%%%%%%%%%%%%%%%%%%%%%%%%%%%%%%%%%%%%%%%%%%%%%%%%%%%%%

\vskip 1cm

\includegraphics{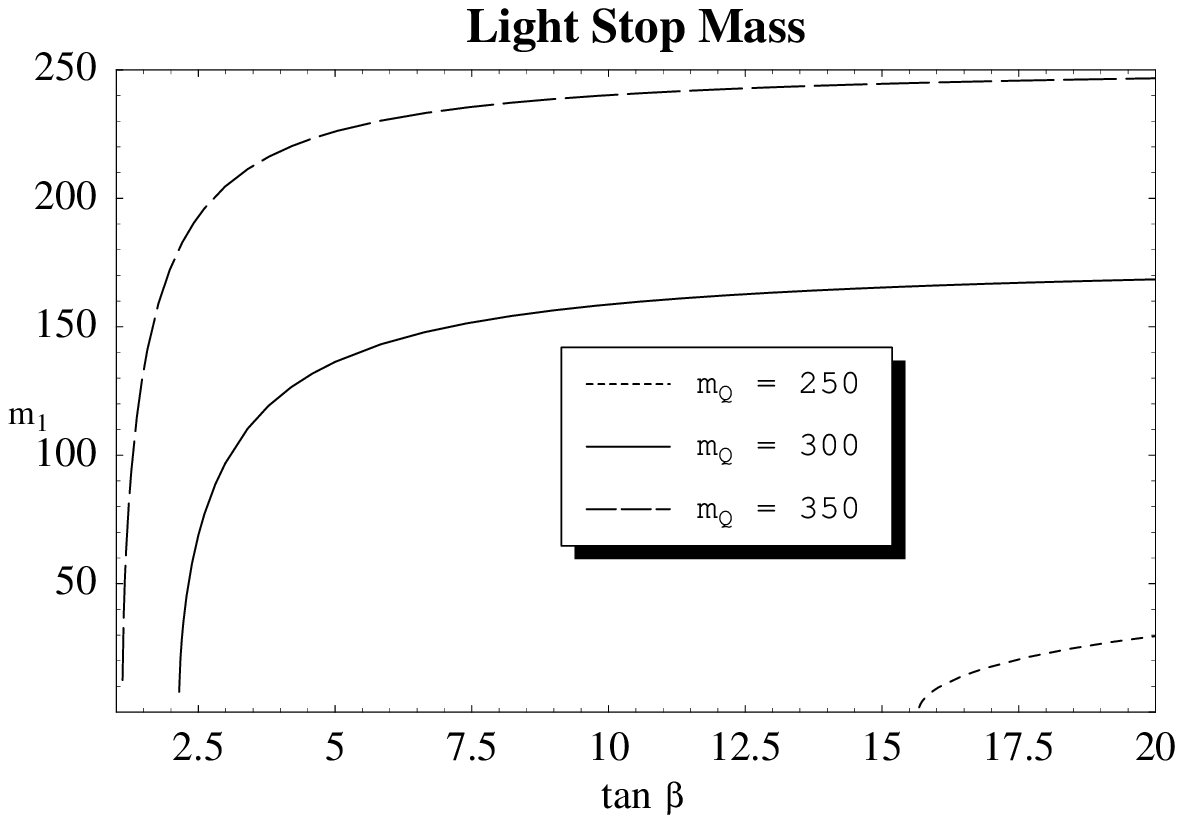}
\vskip 280pt
\begin{itemize}
\item[Fig. 4] Light stop mass $m_1$ in GeV versus $\tan\beta$ 
for the case $m_{Q}=m_{U}= 250,300,350\,$GeV,  
$\mu=500\,$GeV, $A_{t}=500 e^{\frac{i \pi}{4}}\,$GeV.

\newpage\ \quad \vskip -0.5cm

\includegraphics{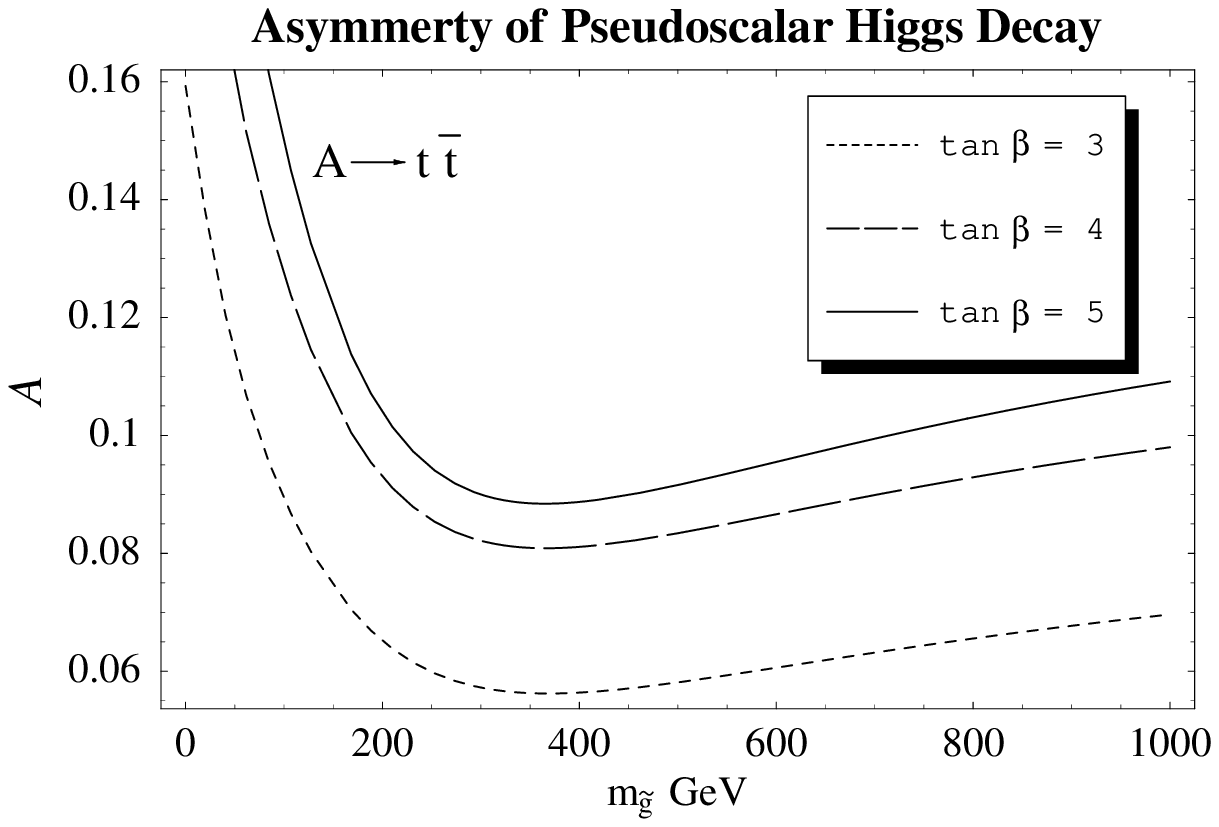}
\vskip 280pt
\item[Fig. 5] Asymmetry of pseudoscalar Higgs decay, 
$m_{Q}=300\,$Gev, $m_{U}=300\,$GeV,  
$\mu=500\,$GeV, $A_{t}=500 e^{\frac{i \pi}{4}}\,$GeV,
$m_{A}=400\,$GeV, $m_{H}=420\,$GeV.

\vskip 1.5cm

\includegraphics{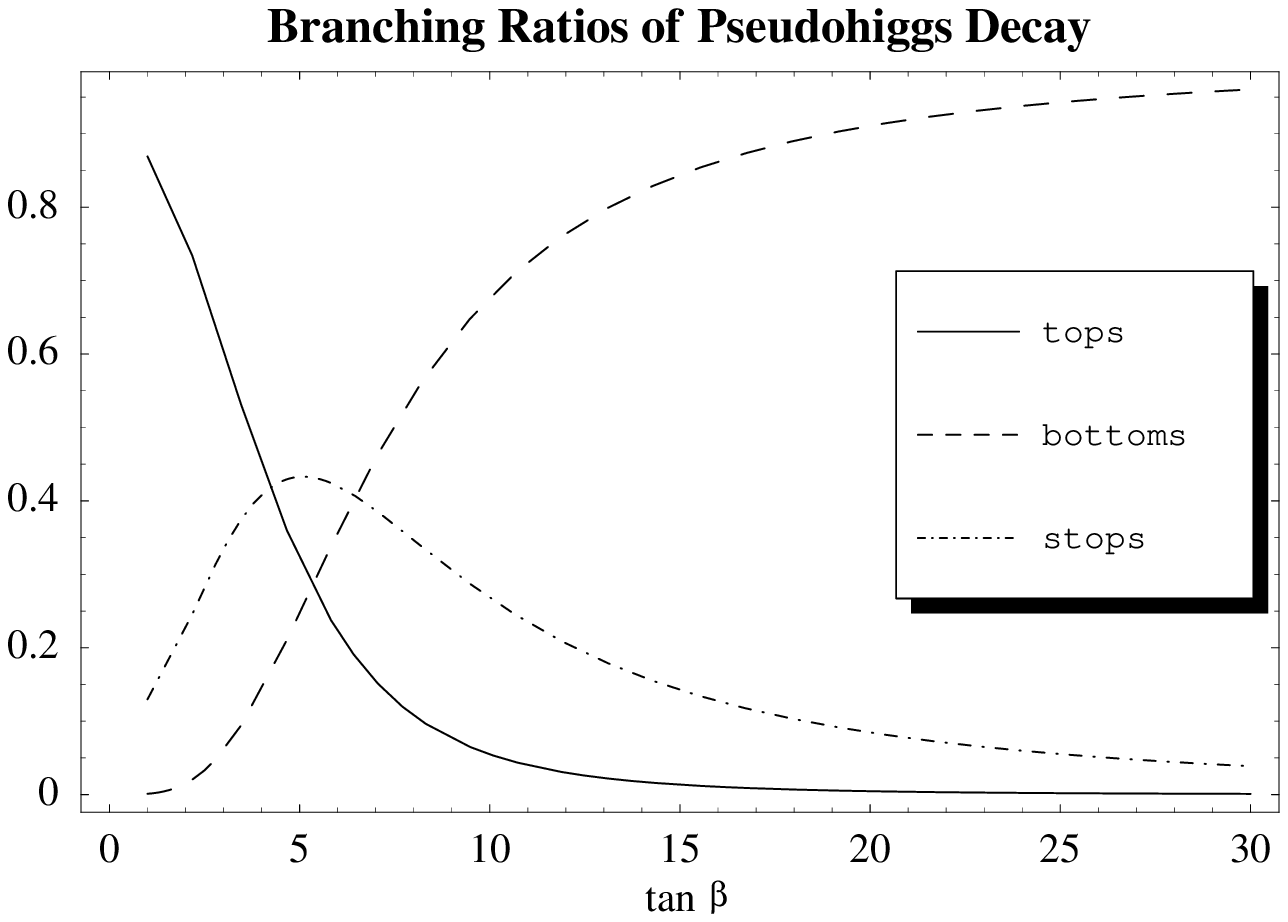}
\vskip 280pt
\item[Fig. 6] Branching ratios for pseudoscalar Higgs decay, 
$m_{Q}=300\,$Gev, $m_{U}=300\,$GeV,  
$\mu=500\,$GeV, $A_{t}=500 e^{\frac{i \pi}{4}}\,$GeV,
$m_{A}=400\,$GeV, $m_{H}=420\,$GeV.

\newpage\ \quad\vskip -0.5cm

\includegraphics{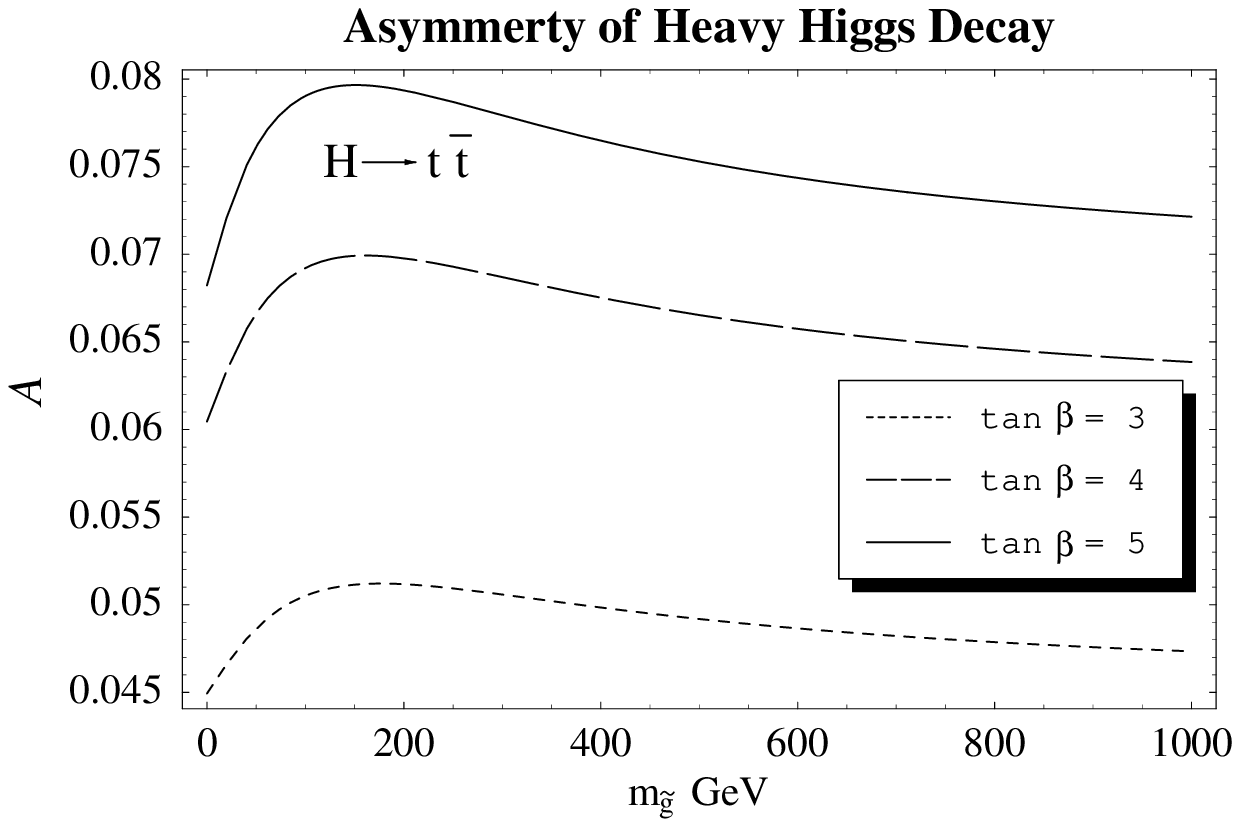}
\vskip 280pt
\item[Fig. 7] Asymmetry of heavy Higgs decay, 
$m_{Q}=300\,$Gev, $m_{U}=300\,$GeV,  
$\mu=500\,$GeV, $A_{t}=500 e^{\frac{i \pi}{4}}\,$GeV,
$m_{A}=400\,$GeV,$m_{H}=420\,$GeV.

\vskip 1.5cm

\includegraphics{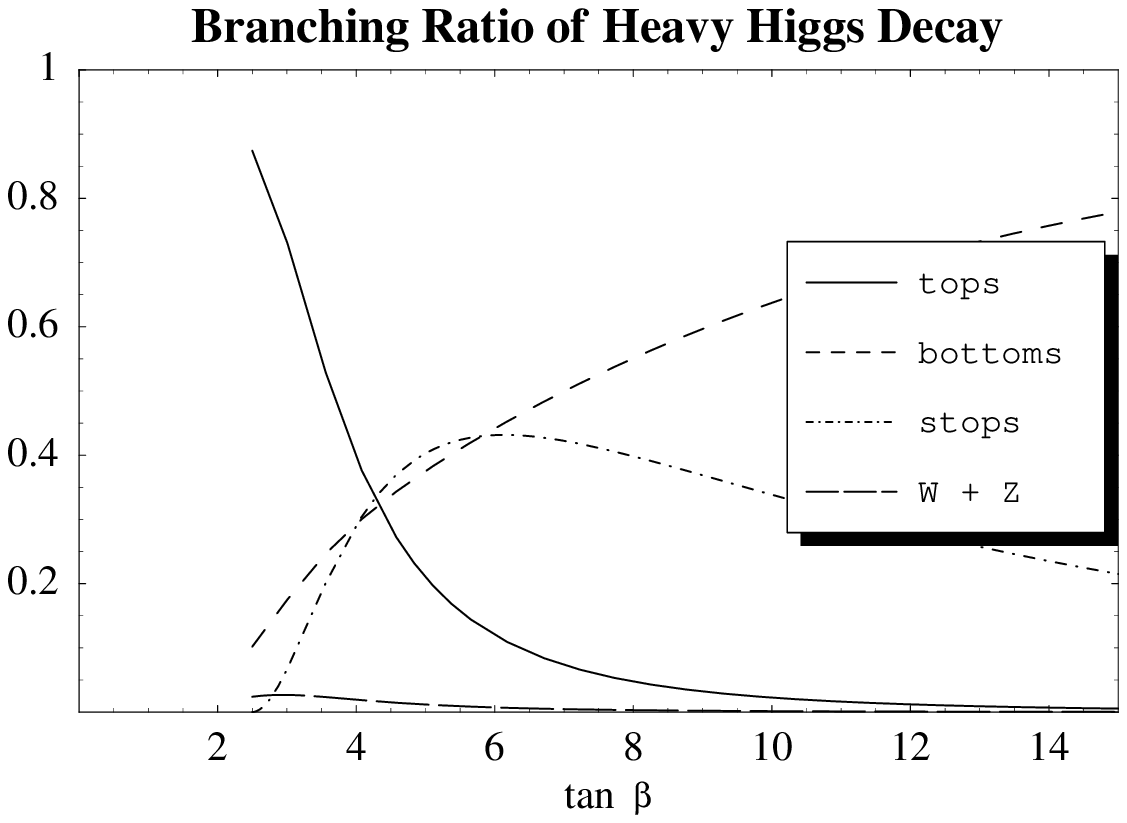}
\vskip 280pt
\item[Fig. 8] Branching ratios for heavy Higgs decay, 
$m_{Q}=300\,$Gev, $m_{U}=300\,$GeV,  
$\mu=500\,$GeV, $A_{t}=500 e^{\frac{i \pi}{4}}\,$GeV,
$m_{A}=400\,$GeV, $m_{H}=420\,$GeV.

\end{itemize}
\end{document}